\documentstyle[preprint,aps]{revtex}

\begin{document}
\draft

\title{Sensitivity curves for spaceborne gravitational wave 
interferometers}

\author{Shane L.\ Larson\cite{Lar} and William A.\ Hiscock\cite{His}}
\address{Department of Physics, Montana State University, Bozeman,\\
Montana 59717}

\author{Ronald W. Hellings\cite{Hel}}
\address{Jet Propulsion Laboratory, Pasadena, CA 91103}

\date{December 07, 1999} \preprint{MSUPHY99.04}

\maketitle

\begin{abstract}
To determine whether particular sources of gravitational radiation 
will be detectable by a specific gravitational wave detector, it is 
necessary to know the sensitivity limits of the instrument.  These 
instrumental sensitivities are often depicted (after averaging over 
source position and polarization) by graphing the minimal values of 
the gravitational wave amplitude detectable by the instrument versus 
the frequency of the gravitational wave.  This paper describes in 
detail how to compute such a sensitivity curve given a set of 
specifications for a spaceborne laser interferometer gravitational 
wave observatory.  Minor errors in the prior literature are corrected, 
and the first (mostly) analytic calculation of the gravitational wave 
transfer function is presented.  Example sensitivity curve 
calculations are presented for the proposed LISA interferometer.  We 
find that previous treatments of LISA have underestimated its 
sensitivity by a factor of $\sqrt{3}$.
\end{abstract}

\pacs{}

\section{INTRODUCTION}\label{sec:intro}

Advances in modern technology have ushered in an era of large laser 
interferometers designed to be used in the detection of gravitational 
radiation, both on the ground and in space.  Such projects include the 
LIGO and VIRGO \cite{LIGO,VIRGO} ground-based interferometers, and the 
proposed LISA and OMEGA \cite{LPPA2,OMEGA} space-based interferometers.  
As these detectors come on-line, a new branch of astronomy will be 
created and a radically new view of the Universe is expected to be 
revealed.  With the era of gravitational wave astronomy on the horizon, 
much effort has been devoted to the problem of categorizing sources of 
gravitational radiation, and extensive studies are underway to 
determine what sources will be visible to the various detectors.

Typically, the sensitivity of detectors to sources of gravitational 
radiation has been illustrated using graphs which compare source 
strengths (dimensionless strain) to instrument noise as functions of 
the gravitational wave frequency.  Many different types of plots have 
appeared in the literature, ranging from single plots of spectral 
density to separate amplitude plots for each class of source.  When 
considering the possibility of observing a new source, or comparing 
aspects of various proposed gravitational wave observatories, it is 
important to be able to generate consistent and accurate noise curves 
for a given instrument and to understand what assumptions have gone 
into generating the curves.  This is especially important in the case 
of spaceborne gravitational observatories which are sensitive to Low 
Frequency (LF) gravitational waves (in the band from 10$^{-4}$ Hz to 1 
Hz).  In this LF band, the sources of radiation include both {\it known} 
continuous sources at well-determined strengths and 
frequencies\cite{WhiteDwarfs}, for which a signal-to-noise ratio only 
slightly greater than one is needed for detection, as well as 
speculative short-lived ``burst'' sources requiring a much greater 
signal-to-noise ratio for detection.

This paper reviews the mathematical formalism and methodology for 
generating noise curves for a class of spaceborne gravitational wave 
interferometers.  A synthesis is provided here of material which has 
hitherto been scattered across the literature, and a variety of new 
results are incorporated.  The gravitational wave transfer function of 
an interferometer, averaged over source direction and polarization, is 
calculated here for the first time,\footnote{While this paper was in 
preparation, we became of aware of recent work by Armstrong, 
Estabrook, and Tinto \cite{AET} which numerically estimates the 
gravitational wave transfer function using Monte Carlo simulations to 
conduct the averaging.  A visual inspection of the sensitivity curve 
derived from the transfer function derived by \cite{AET} shows 
agreement with the results derived in this paper.} and a number of 
minor errors in the existing literature are corrected.  In particular, 
we find that several previous estimates of the sky- and 
polarization-averaged sensitivity of space interferometers have 
understated their sensitivity by a factor of $\sqrt{3}$.  Although the 
results of this paper are applicable to any spaceborne laser 
interferometer system designed for gravitational wave detection, 
specifics of the proposed LISA mission are used as an example.

The outline of the paper is as follows.  In Section 
\ref{sec:instrument}, the concept of instrumental operation for a 
spaceborne interferometer is described.  Section 
\ref{sec:NoiseSpectra} discusses noise sources that limit the 
sensitivity of the detector and Section \ref{sec:SenseCurves} presents 
the mathematical formalism for generating sensitivity curves from 
these noise sources.  Section \ref{sec:Reconcile} considers the 
various types of sensitivity curves which currently exist in the 
literature and reconciles the different methods.

\section{INSTRUMENT OPERATION}\label{sec:instrument}

\subsection{Instrument Design}\label{sub:design}

The common design concept for proposed spaceborne interferometers 
consists of a constellation of probes arranged in an equilateral 
triangle inscribed on the circle of the probes' relative orbits.  In 
the simplest configuration, $3$ probes are used to form a single 
Michelson interferometer, as shown in Figure \ref{ProbeGeometry}.  The 
probe at the vertex of the angle corresponds to the central mirror in 
the interferometer, while the two probes on the ends of the arms 
correspond to the end mirrors.  The effect of a gravitational wave 
passing by the detector is to stretch or contract space in the arms of 
the interferometer.  The effects in the two arms will in general be 
different, due to the different orientation of the two arms in space.  
The wave is detected by monitoring the times of flight of laser 
signals between the probes in order to observe and measure this 
difference.  The basic operation of a space gravitational wave 
detector consists in measuring the phases of the incoming laser 
signals relative to those of the outgoing signals in both arms of the 
interferometer and differencing these relative phases to cancel common 
phase noise in the two arms.

In order to avoid spurious motions of the spacecraft that could mimic 
the effect of a gravitational wave on the laser tracking signals, each 
spacecraft is equipped with a position control system.  This system 
consists of, first, an accelerometer that measures the motion of the 
spacecraft relative to a proof mass that floats freely at the center 
of the accelerometer and, second, a set of thrusters that accelerate 
the spacecraft.  A control loop fires the thrusters in such a way as 
to null the output from the accelerometer and maintain the spacecraft 
on a purely gravitational trajectory.

\subsection{Gravitational Wave Response}\label{sub:gwresponse}

The response of an electromagnetic tracking signal to the passage of a 
gravitational wave has been shown by Estabrook and Wahlquist 
\cite{EstaWahl} to be a Doppler shift in the frequency of the received 
signal relative to the outgoing signal.  For a gravitational wave of 
amplitude $h(t)$, the shift in a signal of fundamental frequency
$\nu_{o}$ will be given by
\begin{eqnarray}
  {{\Delta \nu (t,\theta ,\psi )} \over {\nu_{o}}} &=& {1 \over 2} \cos 
  2\psi  \nonumber \\
  &&\ \times \left[ (1-\cos \theta )h(t)+2\cos \theta ~h(t-\tau -\tau \cos
  \theta )-(1+\cos \theta )h(t-2\tau )\right] ,
  \label{DopplerSignal}
\end{eqnarray}
where $\tau $ is the light travel time between end masses, $\theta $ 
is the angle between the lines-of-sight to the probe and to the 
source, and $\psi $ is a principal polarization angle of the 
quadrupole gravitational wave.

The angles $\theta$ and $\psi$ will, for a general spaceborne 
interferometer, be slowly varying functions of time, dependent upon 
the orbital configuration of the detector.  This paper ignores these 
time dependencies, which are specific to a particular mission design, 
so that the only time dependence assumed in Eq.(\ref{DopplerSignal}) 
is the time varying amplitude of the gravitational wave, $h(t)$.  The 
sky-averaged sensitivity obtained by averaging over the antenna 
pattern associated with a particular interferometer orbital 
configuration will be negligibly different from the all-sky average 
obtained here for a fictitious spatially fixed 
interferometer\footnote{The angles $\theta$ and $\psi$ will vary at 
the orbital period of the interferometer, $T_{orb}$.  In calculating 
the response to a gravitational wave of period $T_{gw}$, the relative 
error in the power spectrum made by assuming the angles to be constant 
will be of order $(T_{gw}/T_{orb})^{2}$.  As long as the orbital 
period is much longer than the gravitational wave period, this error 
will be negligible.}.

It is useful to write $h(t)$ in terms of its Fourier transform 
$\tilde{h}(\omega)$.  If the Doppler record is sampled for a time $T$ 
then $h(t)$ is related to its Fourier transform by
\begin{equation}
    h(t) = {{\sqrt{T}}\over{2 \pi}} \int_{-\infty}^{+\infty} 
    \tilde{h}(\omega) e^{i\omega t} d\omega \ ,
    \label{FourierTransform}
\end{equation}
where the $\sqrt{T}$ normalization factor is used to keep the power 
spectrum roughly independent of time.  Using this definition of the 
Fourier transform, the frequency shift can be written as
\begin{eqnarray}
    \ \Delta \nu (t,\theta ,\psi ) & = & {{\nu_{o}\sqrt{T}}\over{2 \pi}}
    \int_{-\infty}^{+\infty} 
    {1 \over 2} \cos (2 \psi) \ \tilde{h}(\omega ,\theta ,\psi)
    \nonumber \\
    & & \times \left[ (1 - \mu) + 2 \mu e^{-i \omega \tau(1 + \mu)} - 
    (1 + \mu)e^{-i 2 \omega \tau} \right] e^{i \omega t} d\omega \ ,
    \label{DopplerSignal2}
\end{eqnarray}
where $\mu \equiv \cos \theta$.  The quantity that is actually read 
out by the laser interferometer tracking system is phase, so Eq.\ 
(\ref{DopplerSignal2}) is integrated to find the phase in cycles
\begin{equation}
  \Delta \phi (t,\theta ,\psi )=
  \displaystyle \int_{0}^{t} 
  \Delta \nu (t',\theta ,\psi )\text{\ }dt'.
  \label{phase signal}
\end{equation}
If the phase is divided by the laser frequency and the one-way 
light travel time of each arm, one obtains:
\begin{eqnarray}
	z(t,\theta,\psi) & = & {{\Delta \phi (t,\theta ,\psi )} \over 
	{\nu_{o} \tau }} = {{\sqrt{T}}\over{4 \pi \tau}} 
	\int_{-\infty}^{+\infty} d\omega \cos (2 \psi) \ \tilde{h}(\omega) 
	\nonumber \\
    && \times \left[ (1 - \mu) + 2 \mu 
    e^{-i \omega \tau(1 + \mu)} - (1 + \mu)e^{-i 2 \omega \tau}
    \right] {1 \over {i \omega}} e^{i \omega t} \ ,
    \label{strain signal}
\end{eqnarray}
where Eq.\ (\ref{DopplerSignal2}) has been used to expand $\Delta \nu 
(t,\theta ,\psi )$, and arbitrary phases have been set to zero in the 
integration of Eq.\ (\ref{phase signal}). The $1/\omega $ in the 
Fourier integral arises from the integration in the time domain in 
Eq.\ (\ref{phase signal}).  This $z(t,\theta ,\psi )$ has the property 
that, in the limit of low frequency, it reduces to a pure spatial 
strain.  At high frequencies, the signal is much more complicated due 
to the three $h$-terms that enter at different times.  Nevertheless, 
in what follows we will loosely refer to the quantity $z(t,\theta 
,\psi )$ as the gravitational wave strain.

The goal of gravitational wave detection is to detect the strain 
produced by a gravitational wave signal, as given in Eq.\ (\ref{strain 
signal}), in the presence of competing noise.  The primary sources and 
spectra of this noise will be discussed in the next section.

\section{NOISE SPECTRA}\label{sec:NoiseSpectra}

\subsection{Common noise spectra}\label{sub:CommonNoise}

The signal received in a single arm of the interferometer is given by
\begin{equation}
    s_i(t) = z(t,\theta _i,\psi _i) + p(t) - p(t-2\tau _i)
    + n_i(t),  
    \label{s(t)}
\end{equation}
where $s_1(t)$ and $s_2(t)$ are the two noisy strain signals in the 
two arms of the interferometer, $p(t)$ is the laser phase noise which 
is common to the two arms, $n_i(t)$ is the strain noise in the 
$i^{{\rm th}}$ arm produced by all other noise sources, and $\tau_i$ 
is the one-way light travel time in the $i^{{\rm th}}$ arm.  The major 
source of noise in each arm, by several orders of magnitude, is the 
$p(t)$ phase noise in the lasers.  However, this noise is common to 
both arms of the interferometer and can be eliminated through signal 
processing.  When an interferometer signal
\begin{eqnarray}
     \Sigma(t) &=& s_{1}(t) - s_{2}(t) \nonumber \\
               &=& z_{1}(t) - z_{2}(t) + 
             n_{1}(t) - n_{2}(t) - p(t-2\tau_1) + p(t-2\tau_2)
 \label{convSignal}
\end{eqnarray}
is formed, the laser phase noise will cancel as $\tau_1 \rightarrow 
\tau_2$.  Unfortunately, a space-based interferometer consisting of 
freely flying spacecraft will necessarily have unequal armlengths, 
preventing effective use of $\Sigma(t)$ as a signal for data analysis.

However, a new data reduction procedure for space interferometry has been
recently discovered and discussed in a paper by Tinto and Armstrong
\cite{tinto}, that eliminates all laser phase noise even if the two
light travel times $\tau_1$ and $\tau_2$ are unequal. Working in the
time domain, one defines the combination
\begin{eqnarray}
   X (t) &=& s_1(t) - s_2(t) - s_1(t-2\tau _2) + s_2(t-2\tau _1) 
   \nonumber \\
   &=& z_{1}(t) - z_{2}(t) - z_{1}(t - 2\tau_{2}) + z_{2}(t - 
   2\tau_{1})  \nonumber \\
   &&\ \ \ + n_{1}(t) - n_{1}(t - 2\tau_{2}) - n_{2}(t) + n_{2}(t - 
   2\tau_{1})\ ,
\label{Sigma final}
\end{eqnarray}
which is devoid of laser phase noise for all values of the two light 
travel times $\tau_1$ and $\tau_2$.  In order for the laser phase 
noise to cancel exactly, both light travel times must be known exactly 
so that the combination of signals in Eq.\ (\ref{Sigma final}) can be 
correctly formed.  In practice, however, all that is necessary is for 
the armlengths to be known well enough for the laser phase noise to 
become insignificant relative to the independent noise sources 
(discussed next).

\subsection{Independent noise spectra}\label{sub:IndependentNoise}

Since noise common to the two arms of an interferometer can be 
significantly reduced by the procedure just described, the limitation 
of sensitivity of the space-based gravitational wave missions is 
actually determined by noise sources that are independent in the two 
arms.  There are two major types of such independent noise: position 
noise in the laser tracking system and acceleration noise due to 
parasitic forces on the proof mass of the accelerometer.

Position readout noise is noise that acts like a change in the length 
of the optical path for the tracking signal.  The strain spectral 
density it creates is given by
\begin{equation}
  S_n(f)=\frac{S_x}{c^2\tau ^2}
  \label{position noise}
\end{equation}
where $S_x$ is the spectral density of position noise in 
m$^2\,$Hz$^{-1}$.  The most important single source of position noise 
in the current designs is shot noise in the lasers. This may therefore
be used as a rough indicator of the expected position noise for a
particular interferometer.

Shot noise produces a phase noise given by
\begin{equation}
   S_\phi=\frac{h\nu_{o}}{P_r}
   \label{shot noise defn}
\end{equation}
where $h\nu_{o}$ is the photon energy and $P_{r}$ is the received power.  
The received power is calculated from
\begin{equation}
		 P_r=P_t\left[{\epsilon \pi^2 \nu_{o}^2 D^2 \over
		 c^2} \right] 
		 \left[\frac{1}{4\pi r^2}\right] \left[\frac{\epsilon\pi 
		 D^2}{4}\right]
  \label{received power}
\end{equation}
where $P_t$ is the transmitted power , the quantity in the first 
square brackets is the directional gain of the transmitter optics with 
diameter $D$ and efficiency $\epsilon$ for light of frequency $\nu_{o}$,
the quantity in the second square brackets is the space 
loss at a distance $r$, and the quantity in the last square brackets 
is the effective cross section of the receiving optics.  The effective 
position noise produced by this phase noise is
\begin{equation}
	S_{x} = {c^2\over{4\pi^2\nu_{o}^2}}S_\phi.
	\label{shot position}
\end{equation}
Combining Eqs.\ (\ref{position noise}-\ref{shot position}), the 
formula for the 
strain noise produced by shot noise in the lasers is
\begin{equation}
	S_n = {4c^4 \over{\pi ^4}}{{h\nu_{o}} \over {\epsilon^{2}D^{4} 
	P_{t}}} {1 \over {\nu_{o}^{4}}}\ ,
	\label{ShotNoise}
\end{equation}
where we have used the fact that $r\approx c\tau$. 

A more detailed study will include other, smaller sources of position noise.
For the current LISA design, the strengths of many of these potential 
additional noise sources have been budgeted. The LISA shot noise
is expected to have spectral density $S_x(f)=1.2\times10^{-22}$ 
m$^2\,$Hz$^{-1}$ , while the total position noise is estimated to have 
spectral density $S_x(f)=1.6\times10^{-21}$ m$^2\,$Hz$^{-1}$ \cite{LPPA2}.

In addition to position noise, there will be acceleration noise in the
detectors. Acceleration noise produces a strain noise spectral 
density given by
\begin{equation}
  S_n(f)=\frac{S_a}{(2\pi f)^4(c\tau )^2}
  \label{AccelerationNoise}
\end{equation}
where $S_a$ is the acceleration spectral density in 
m$^2$s$^{-4}$Hz$^{-1}$.  The major source of acceleration noise in the 
space detectors is expected to be parasitic forces acting on the proof 
mass of the accelerometer.  These instruments are very complex and the 
noise is difficult to characterize, especially in the laboratory tests 
where the proof mass must be suspended in one-$g$ rather than being 
allowed to float freely in space.  Nevertheless, from what has been 
learned in the laboratory, it appears that a flat acceleration noise 
spectrum at a level of $S_a = 9\times10^{-30}$\ m$^2$\ s$^{-4}$\ 
Hz$^{-1}$ should be achievable over most of the frequency band of 
interest (this is the level assumed in the present LISA design 
\cite{LPPA2}).

The total noise curves for space gravitational wave detectors are 
found by adding the various noise spectra.  Using the values for 
$S_x(f)$ and $S_a(f)$ given above, a total spectral density of strain
noise in the detectors can be found by adding Eqs.\ (\ref{position noise}) 
and (\ref{AccelerationNoise}) in quadrature.  The resulting root 
spectral density of the instrument noise is plotted in Figure 
\ref{NoiseSpectrum}. The quoted LISA spectral densities have been used
for the acceleration and total position noises: $S_a = 9\times 10^{-30}$
\ m$^2$\ s$^{-4}$\ Hz$^{-1}$ and $S_x(f) = 1.6 \times 10^{-21}$ 
m$^2\,$Hz$^{-1}$.

\section{GENERATION OF SENSITIVITY CURVES}\label{sec:SenseCurves}

The sensitivity of an instrument to a gravitational wave depends on the
relationship between the amplitude of the wave and the size of the signal
that eventually appears in the detector.  It also depends on the size of 
the noise in the final output signal of the instrument.  The connection
between amplitude and signal is calculated in frequency space and is 
called the {\it transfer function}.  If the interferometer output
$\Sigma(t)$ has a gravitational wave contribution to it given by
\begin{equation}
  \Delta(t) = z_{1}(t) - z_{2}(t) \ ,
  \label{DeltaDef}
\end{equation}
then the transfer function  $R(\omega)$ is defined by
\begin{equation}
   S_{\bar{\Delta}}(\omega) = S_{h}(\omega) R\left(\omega\right) \ ,
   \label{SREquation}
\end{equation}
where the gravitational wave amplitude spectral density $S_{h}(\omega)$ 
is defined by
\begin{equation}
	S_h(\omega) = | \tilde{h}(\omega) |^2 \ ,
\label{SpectralDefinition}
\end{equation}
so that the mean-square gravitational wave strain is given by
\begin{equation}
   \langle h^{2} \rangle = {1 \over T} \int_{0}^{\infty} h (t)^{2} 
   dt = {1 \over {2 \pi}} \int_{0}^{\infty} S_{h}(\omega) d\omega \ .
   \label{SpectralDefinitionProps}
\end{equation}
Similarly, the instrumental response $S_{\bar{\Delta}}(\omega)$ is 
defined such that
\begin{equation}
   \overline{\langle \Delta^{2} \rangle}
   = {1 \over {2 \pi}} \int_{0}^{\infty} S_{\bar{\Delta}}(\omega)
   d\omega \ ,
   \label{SpectralPower}
\end{equation}
where the brackets indicate a time average and the bar over the 
$\Delta$ in Eq.\ (\ref{SpectralPower}) indicates that it is averaged 
over source polarization and direction.

In the next section, the transfer function from the gravitational wave 
amplitude $h$ to the interferometer signal $\bar\Delta$ (which is the 
signal part of $\Sigma$) is worked out.  As discussed in Section 
\ref{sub:CommonNoise}, the preferred instrumental output would 
actually be $X(t)$, defined in Eq.\ (\ref{Sigma final}), since it 
exactly cancels the laser phase noise in the data.  However, $X(t)$ is 
a difficult quantity to work with since its transfer function will 
depend on the particular values of the interferometer arm lengths, 
$\tau_{1}$ and $\tau_{2}$.  One approach to simplify the calculation 
is to assume $\tau_{1} = \tau_{2}$ (This approach is taken in Ref.\ 
\cite{AET}.)  At the end of the next section, we will give the 
transfer function for $X(t)$ with this assumption made, and will 
show that in this limit $\Sigma(t)$ and $X(t)$ yield the same
instrumental sensitivity.

\subsection{Gravitational Wave Transfer Function}\label{sub:Transfer}

Previous estimates of sensitivity \cite{HelBlair} have often 
resorted to working in the long wavelength approximation, beginning 
from the Doppler tracking signal described by Eq.\ 
(\ref{DopplerSignal}), or have combined results for independent single 
arms.  Here the {\it exact} gravitational wave transfer function is 
computed, without such approximations, from the gravitational wave 
strain.

Consider the geometry shown in Figure \ref{GeometryFig}.  The vectors 
$\bf{L}_{1}$ and $\bf{L}_{2}$ point along the arms of the 
interferometer, and the vector $\bf{k}$ points along the propagation 
vector of the gravitational wave.  The quantities $\theta_{i}$ measure 
the angular separation between the arm vectors and the propagation 
vector.  The value $\gamma$ is the opening angle of the 
interferometer, and $\epsilon$ is the inclination of the plane 
containing $\bf{k}$ and $\bf{L_{1}}$ to the plane of the 
interferometer.

Not shown are the principal polarization vectors of the gravitational 
wave, which lie in a plane $90$ degrees away from ${\bf k}$.  The 
polarization angles, $\psi_{i}$, are measured from the point where the 
plane of the $i^{th}$ arm and the propagation vector intersects the 
plane containing the principal polarization vector.  The angle 
$\alpha$ in Figure \ref{GeometryFig} is simply the difference of the 
two polarization angles, $\alpha = \psi_{2} - \psi_{1}$.

The average power in the interferometer is given by
\begin{equation}
\langle \Delta^{2} \rangle = \lim_{T \rightarrow \infty} {1 \over T}
\int_{0}^{\infty} \left| \Delta \right|^{2} dt \ ,
\label{PowerDef}
\end{equation}
where $\Delta$ is defined by Eq.\ (\ref{DeltaDef}).  Using the 
definition of $z$ from Eq.\ (\ref{strain signal}) this can be expanded 
to yield
\begin{equation}
  \langle \Delta^{2} \rangle = {1 \over {2 \pi}}
      \int_{0}^{\infty} d\omega \ \tilde{h}^{2}(\omega) {1 
      \over (\omega \tau)^{2}} \left[ T_{1}(\omega) + T_{2}(\omega) - 2 
      T_{3}(\omega) \right] \ ,
  \label{Power}
\end{equation}
where
\begin{eqnarray}
  T_{1}(\omega) & = & \cos^{2}(2 \psi_{1}) \left[ \mu_{1}^{2}\left(1 + 
  \cos^{2}(\omega \tau) \right) + \sin^{2}(\omega \tau) \right. \nonumber \\
  & - & \left. 2 \mu_{1}^{2} \cos (\omega \tau) \ \cos (\omega \tau \mu_{1})
  - 2 \mu_{1} \sin (\omega \tau) \ \sin (\omega \tau \mu_{1}) \right] \ ,
  \label{T1} \\
  T_{2}(\omega) & = & \cos^{2}(2 \psi_{2}) \left[ \mu_{2}^{2}\left(1 + 
  \cos^{2}(\omega \tau) \right) + \sin^{2}(\omega \tau) \right. \nonumber \\
  & - & \left. 2 \mu_{2}^{2} \cos (\omega \tau) \ \cos (\omega \tau \mu_{2})
  - 2 \mu_{2} \sin (\omega \tau) \ \sin (\omega \tau \mu_{2}) \right] \ ,
  \label{T2} \\
  T_{3}(\omega) & = & \cos (2 \psi_{1}) \ \cos (2 \psi_{2}) \ \eta(\omega)  \ , 
  \label{T3}
\end{eqnarray}
with $\mu_{i} = \cos \theta_{i}$, and where
\begin{eqnarray}
   \eta(\omega, \theta_{1}, \theta_{2}) = \left[ \cos (\omega
   \tau) - \cos (\omega \tau \mu_{1}) \right]
   \left[ \cos (\omega \tau) - \cos (\omega \tau \mu_{2})
   \right] \mu_{1} \mu_{2} \nonumber \\
   + \left[ \sin (\omega \tau) - \mu_{1} \sin
   (\omega \tau \mu_{1}) \right] \left[ \sin (\omega \tau) -
   \mu_{2} \sin (\omega \tau \mu_{2}) \right] \ ,
   \label{eta}
\end{eqnarray}
has been defined for convenience.  The expression for the power in the 
detector, as given by Eq.\ (\ref{Power}), is a complicated function of 
frequency and of the orientation between the propagation vector of the 
gravitational wave and the interferometer.  It represents the antenna 
pattern for a laser interferometer gravitational wave detector.  At 
low frequencies, the frequency dependence drops out and the 
expressions simplify greatly \cite{Giampieri}, but at 
higher frequencies it remains a very complicated frequency-dependent 
object.

To characterize the average sensitivity of the instrument it is 
customary to consider the isotropic power, obtained by averaging the 
antenna pattern over all propagation vectors and all polarizations:
\begin{equation}
  \overline{\langle \Delta^{2} \rangle} = {1 \over {8 \pi^{2}}}
  \int_{0}^{2 \pi} d\psi \ 
  \int_{0}^{2 \pi} d\epsilon \ 
  \int_{0}^{\pi} \sin \theta \ d\theta \ \langle \Delta^{2} \rangle \ .
  \label{AvgPower}
\end{equation}

Since the variables $\left(\theta_{1}, \psi_{1}\right)$ and 
$\left(\theta_{2}, \psi_{2}\right)$ that figure in Eq.\ (\ref{Power}) 
each locate the same propagation vector of the gravitational wave, 
they are not independent of one another.  However, examination of Eqs.  
(\ref{T1}-\ref{T3}) shows that $T_{1}$ depends only on the 
$\left(\theta_{1}, \psi_{1}\right)$ variables, while $T_{2}$ depends 
only on $\left(\theta_{2}, \psi_{2}\right)$.  This allows the 
integration of Eq.\ (\ref{AvgPower}) to be performed very easily for 
the $T_{1}$ and $T_{2}$ terms without converting to a common set of 
angular variables.  In each case, the averaging of the $T_{1}$ and 
$T_{2}$ terms over the $\left(\theta_{i}, \psi_{i}\right)$ gives 
precisely the same value:
\begin{eqnarray}
   \overline{T_{1}} & = & \overline{T_{2}} =
   {1 \over {8 \pi^{2}}} \int d\psi_{1} \ d\epsilon \  d\theta_{1} \ 
   \sin \theta_{1} \ T_{1} \nonumber \\
   & = &
   {1 \over 2} \left[ \left(1 + \cos^{2} (\omega \tau) \right) \left({1 
   \over 3} - {2 \over{(\omega \tau)^{2}}} \right) + \sin^{2} (\omega 
   \tau) + {4 \over{(\omega \tau)^{3}}} \sin (\omega \tau) \ \cos (\omega 
   \tau) \right] \ .
   \label{T1Avg}
\end{eqnarray}
The average isotropic power can then be expressed as
\begin{equation}
     \overline{\langle \Delta^{2} \rangle} =
     {1 \over \pi} \int_{0}^{\infty} d\omega \ \tilde{h}^{2}(\omega)
     {1 \over (\omega \tau)^{2}} \left[ \overline{T_{1}} -
     \overline{T_{3}} \right] \ .
   \label{isotropicPower}
\end{equation}

To complete the integration, the function $T_{3}$ (which depends on 
both $\left(\theta_{1}, \psi_{1}\right)$ and $\left(\theta_{2}, 
\psi_{2}\right)$) must finally be expressed in terms of a single set 
of angular variables.  One may choose to eliminate $\left(\theta_{2}, 
\psi_{2}\right)$ in favor of $\left(\theta_{1}, \psi_{1}\right)$ by 
using conventional spherical trigonometry in Figure \ref{GeometryFig}.  
For the polarization angle, the relationship is particularly simple
\[
  \psi_{2} = \psi_{1} + \alpha
\]
and the integration over $\psi_{1}$ may be carried out analytically, 
giving,
\begin{equation}
  \overline{T_{3}} = {1 \over {8 \pi^{2}}} \int d\psi_{1} \ d\epsilon
  \  d\theta_{1} \ \sin \theta_{1} \ T_{3}
  = {1 \over {8 \pi}} \int d\epsilon \ d\theta_{1} \  \sin \theta_{1}
  \ \cos (2\alpha) \ \eta(\omega,\theta_{1},\theta_{2}) \ .
   \label{T3Avg}
\end{equation}
The cost of carrying out the $\psi_{1}$ integration is the 
introduction of the angle $\alpha$, which can be related to the 
$\{ \epsilon, \theta_{i} \}$ variables using the law of sines 
in Figure \ref{GeometryFig}:
\begin{equation}
	\sin \alpha = {{\sin \gamma \sin \epsilon}\over{\sin \theta_{2}}}
	\ .
    \label{SinAlpha}
\end{equation}
The function $\eta(\omega,\theta_{1},\theta_{2})$ in Eq.\ (\ref{eta}) 
has terms containing $\mu_{2} = \cos \theta_{2}$, which must be 
re-expressed in terms of the integration variable $\theta_{1}$.  The 
relationship between $\theta_{1}$ and $\theta_{2}$ is given by
\begin{equation}
   \cos \theta_{2} = \cos \gamma \ \cos \theta_{1} + \sin \gamma \ \sin 
   \theta_{1} \ \cos \epsilon \ ,
   \label{CosTheta2}
\end{equation}
where $\gamma$ is the opening angle of the interferometer, and 
$\epsilon$ is the inclination of the gravitational wave propagation 
vector to the interferometer.  Due to the complexity of 
$\eta(\omega)$ when Eq.\ (\ref{CosTheta2}) is substituted into Eq.\ 
(\ref{eta}), we have not been able to calculate an expression for 
$\overline{T_{3}}$ analytically, so it will be kept as an explicit 
integral.

Using the definition of $R(\omega)$ from Eq.\ (\ref{SREquation}) with 
the average isotropic power in Eq.\ (\ref{isotropicPower}), the 
gravitational wave transfer function is found to be
\begin{eqnarray}
   R(\omega) & = & 2 {1 \over (\omega \tau)^{2}} \left[
   \overline{T_{1}} - \overline{T_{3}} \right]
   \nonumber \\
   & = & {1 \over (\omega \tau)^{2}}
   \left[ \left(1 + \cos^{2} (\omega \tau) \right) \left({1 
   \over 3} - {2 \over{(\omega \tau)^{2}}} \right) \right. \nonumber \\
   & + & \sin^{2} (\omega \tau) + {4 \over{(\omega \tau)^{3}}}
   \sin (\omega \tau) \ \cos (\omega \tau) \nonumber \\
   & - & \left. {1 \over {4 \pi}} \int d\epsilon \ d\theta_{1} \ 
    \sin \theta_{1}\ \left(1 - 2 \sin^{2} \alpha \right) \eta(\omega)
   \right] \ .
\label{TransferFunction}
\end{eqnarray}
It is straightforward to evaluate the remaining integral using simple 
numerical techniques.  The exact transfer function, including the 
numerical evaluation of the last integral, is shown in Figure 
\ref{TransferPlot} as a function of the dimensionless quantity $u = 
\omega \tau$.

The high frequency structure in Figure \ref{TransferPlot} dominates 
the shape of the transfer function at frequencies greater than 
$\omega \simeq 1/\tau$.  At these frequencies, the armlength $\tau$ of 
the interferometer becomes comparable to the wavelength of 
the gravitational wave.  The extrema in the transfer function are 
amplifications due to interference of the signals in the arms.  The 
minima occur at frequencies
\begin{equation}
    f = n/2 \tau \ .
    \label{Minima}
\end{equation}
The appearance of these periodic amplifications is familiar from basic 
interferometry.

Taking the limit of small $\omega \tau$ in Eq.\ 
(\ref{TransferFunction}) will yield a low frequency limit for 
$S_{\bar{\Delta}}$ of
\begin{equation}
   S_{\bar{\Delta}} = {4 \over 5} \sin^{2}\gamma \:
   S_{h} \ ,
\label{LowFreqLimit}
\end{equation}
which is in agreement with a previous result from Hellings 
\cite{HelBlair}\footnote{In the notation of this paper, the 
interferometer signal has been multiplied by the laser period and 
divided by the arm length of the interferometer, as described in Eq.\ 
(\ref{strain signal}).  To agree with the expressions for the spectral 
density of the interferometer signal $S_{\delta}$ in the literature, 
this factor must be accounted for above, so that $S_{\delta} = \left( 
{\tau \nu_{o}} \right)^{2} S_{\bar{\Delta}}$.} .

\subsection{Sensitivity Curves}

The sensitivity curve for a gravitational wave observatory is obtained 
from Eq.\ (\ref{SREquation}) in the case where the spectral density of 
the isotropic power, $S_{\bar{\Delta}}$, is equal to the spectral 
density of noise in the detector.  The noise in the final instrumental 
signal may be related to the hardware noise in each detector by 
inspection of Eq.\ (\ref{convSignal}).  Assuming that $n_1(t)$ and 
$n_2(t)$ are uncorrelated, the spectral density of the signal produced 
by phase measurement noise is
\begin{equation}
   S_{N} = 4 S_{n}\, ,
   \label{NoiseSpectralPower}
\end{equation}
where $S_n$ is the spectral density of noise in a one-way measurement.  
One factor of $2$ in Eq.\ (\ref{NoiseSpectralPower}) comes from the 
two $n_i(t)$ in Eq.  (\ref{convSignal}) adding in quadrature, since 
they are uncorrelated, and one factor of $2$ arises from the fact that 
the two-way noise in each arm is $\sqrt{2}$ times the one-way noise.  
The final equation for gravitational wave amplitude sensitivity is 
thus
\begin{equation}
   S_{h} = {S_{\bar{\Delta}} \over R} = {{S_{N}} \over R} = {{4 S_{n}} 
   \over R} \ .
   \label{NoiseSpectralPower2}
\end{equation}
The spectral amplitude sensitivity is simply the square root of $S_h$, 
or
\begin{equation}
   h_{f} = \sqrt{S_{h}} = 2 \sqrt{{S_{n} \over R}} \ .
   \label{SpectralAmplitude}
\end{equation}

Using the noise curve given in Figure \ref{NoiseSpectrum}, and using 
the LISA value of $c\tau = 5 \times 10^{9}$ m, the LISA sensitivity 
curve, computed using Eq.\ (\ref{SpectralAmplitude}), is shown in 
Figure \ref{NoiseCurves}.  The position of the high frequency ``knee'' 
occurs at $f = 1/(2 \pi \tau) = 10^{-2}$ Hz.

The greatest sensitivity is seen to occur in a mid-frequency 
``floor''; the level of this floor is set by the size of the position 
noise.  The width of the floor is a function of the acceleration noise 
level and the arm length of the interferometer.  The low frequency 
rise occurs when acceleration noise begins to dominate over position 
noise; the high frequency rise is caused by the turnover in the 
transfer function, at $f \simeq 1/(2 \pi \tau)$.

\subsection{Comparison with Prior Results}\label{sec:Compare}

One may compare the transfer function of Eq.\ (\ref{TransferFunction}) 
with others that have previously appeared in the literature.  These 
transfer functions differ from the one presented in this work in two 
ways.

First, previous results in the literature have approximated the 
transfer function either by working in the low frequency limit, where 
the transfer function becomes constant\cite{HelBlair}\footnote{The 
full transfer function has not been needed for the ground-based 
interferometers such as LIGO since in their operation (as Fabry-Perot 
cavities) the low frequency limit is valid up to about $40-50$ kHz.}, 
or by combining single arm results rather than treating the 
interferometer as a single instrument\cite{schilling}.  In the latter 
cases, the transfer functions do not include the interference term 
between the two arms of the interferometer (the term denoted 
$\bar{T_{3}}$ in Section \ref{sub:Transfer}).

Second, the transfer functions are often incorrectly 
normalized\cite{LPPA2,schilling}.  The problem is that the transfer 
function between gravitational wave amplitude and detector response is 
a well-defined quantity (see Eq.\ (\ref{SREquation})), which is 
inappropriate to renormalize.  When averaged over position and 
polarization, the value of $R(\omega)$ is given by Eq.\ 
(\ref{TransferFunction}).  In the low frequency limit this becomes 
$R(\omega) = (4/5)\sin^{2} \gamma = 3/5$ (Eq.\ (\ref{LowFreqLimit}) 
with $\gamma = 60^{\circ}$).  However, the transfer function as given 
in Refs.\ \cite{LPPA2,schilling} has been normalized by dividing by 
the maximum sensitivity of the instrument, {\it i.e.}, by the response 
for a source with optimal polarization located at a point normal to 
the plane of the detector.  As may be seen by substituting $\mu_i=0$ 
in Eqs.\ (\ref{Power}-\ref{eta}), this maximum instrument response is 
$\langle \Delta^2 \rangle=3 \langle h^2 \rangle$, so division by this 
quantity would give a transfer function whose low-frequency limit is 
$R = 1/5$ rather than $R = 3/5$.  The previous LISA sensitivity curves
thus underestimate the strength of the signal that a gravitational wave
of amplitude $h$ will produce in the detector (see Eq. (16))
and conclude that the minimum detectable amplitude is
\begin{equation}
    h_{\rm{LISA}}= \sqrt{3} h\ ,
\label{LisaH}
\end{equation}
thereby underestimating the sensitivity of the detector.

\subsection{The Transfer Function for $X(t)$}

As discussed at the beginning of Section \ref{sec:SenseCurves}, the
preferred signal from the interferometer for purposes of data analysis
is not $\Sigma(t)$, but $X(t)$, since the common laser phase noise 
exactly cancels in $X(t)$. Unfortunately, the transfer function for
$X(t)$ depends in a complicated way on the particular values of 
$\tau_1$ and $\tau_2$.  However, substantial simplification occurs
in the special case $\tau_1\approx\tau_2=\tau$, which was treated in 
Ref.\ \cite{AET}. In this case, the formula for $X(t)$ becomes
\begin{eqnarray}
   X(t) &=& z_{1}(t) - z_{2}(t) - \left[ z_{1}(t - 2\tau) - z_{2}(t - 
   2\tau) \right] \nonumber \\
   &&\ \ \ + n_{1}(t) - n_{2}(t)
   - \left[ n_{1}(t - 2\tau) - n_{2}(t - 2\tau)\right]\ ,
   \label{equalSigma}
\end{eqnarray}
which may also be written as
\begin{equation}
      X(t) = \Xi(t) + \sigma(t)
\label{NewX}
\end{equation}
where 
\begin{equation}
	\Xi(t) = \Delta(t) - \Delta(t-2\tau) \ ,
\label{xsig}
\end{equation}
and
\begin{equation}
	\sigma(t) = N(t) - N(t-2\tau)\  .
\label{xnoise}
\end{equation}
Here $\Delta(t)$ is given by Eq.\ (\ref{DeltaDef}) and $N(t)\equiv 
n_1(t)-n_2(t)$. 

The portion of $X(t)$ which is a gravitational wave signal is then, by 
Eq.(\ref{xsig}), simply two copies of the $\Delta$ signal we have 
previously analyzed.  The transfer function for $\Xi(t)$ may be 
calculated following the procedure in Sec.[\ref{sub:Transfer}], to 
find
\begin{eqnarray}
	S_{\Xi}(\omega) &=& 4 \sin^{2}(\omega \tau) 
	S_{\bar{\Delta}}(\omega) \nonumber \\
	&=& 4 \sin^{2}(\omega \tau)	S_h(\omega)R(\omega)\ ,
	 \label{TransferFactor}
\end{eqnarray}
where Eq.(\ref{SREquation}) has been used for the second equality.
The transfer function for $X$ will then be
\begin{equation}
	R_{X}(\omega) = 4 \sin^{2}(\omega \tau)R(\omega)\ ,
\label{XTransferFunction}
\end{equation}
where $R(\omega)$ is the previously derived transfer function for 
$\Delta$, given by Eq.\ (\ref{TransferFunction}).

At first glance, it appears the that transfer function for $X$ could 
be as large as four times the transfer function for $\Delta $, and 
hence that weaker gravitational waves could be observed by 
constructing $X$.  However, to determine the sensitivity of the 
interferometer, one must equate $S_{\Xi}(\omega)$ in Eq.\ 
(\ref{TransferFactor}) to the appropriate spectral density of noise.  
For $X$, this is formed using the $\sigma(t)$ combination defined 
in Eq.\ (\ref{xnoise}).  Because the noise contribution to $X$ is 
formed by the same subtraction process, it will similarly have
\begin{equation} 
	S_{\sigma}(\omega)=4 \sin^{2}(\omega \tau)S_N(\omega)\ .
\label{TransferNoise}
\end{equation}	
The additional factors due to the subtraction process used to form 
$X$, which appear in Eqs.\ (\ref{TransferFactor},\ref{TransferNoise}), 
will cancel in the computation of the sensitivity limit of the 
interferometer:
\begin{eqnarray}
	S_h &=& {S_{\sigma} \over R_X} \nonumber \\
	&=&	{{4 \sin^{2}(\omega \tau)S_N(\omega)} \over R_X} =
	{4 S_n \over R}\ ,
\label{Xsens}
\end{eqnarray}
where the second equality follows from Eq.(\ref{XTransferFunction})
and the definition of $N$ in terms of the $n_i$.

Thus, in the limit as $\tau_1 \rightarrow \tau_2$, the sensitivity curve
for $X(t)$ will be identical to that previously computed for $\Delta (t)$,
as shown in Figure \ref{NoiseCurves}.  We are investigating transfer 
functions for unequal arm cases and intend to present these in a future 
paper.

\section{DISCUSSION AND RECONCILIATION}\label{sec:Reconcile}

The final goal of producing noise curves and response functions is to 
answer the question of what gravitational wave sources are detectable.  
Several different approaches have been used in the literature to 
answer this question.  It is the purpose of this section to discuss 
these different approaches and to try to reconcile them.

Section \ref{sec:NoiseSpectra} discussed the actual strain noise 
spectral density ($S_n(f)$, see Figure \ref{NoiseSpectrum}), and 
Section \ref{sec:SenseCurves} showed how the gravitational wave 
transfer function ($R(f)$, Figure \ref{TransferPlot}) is used to 
determine the gravitational wave sensitivity curve ($S_h(f)$, Figure 
\ref{NoiseCurves}).  The gravitational wave transfer function 
actually represents an averaged response of the interferometer,
averaged over waves coming from different directions and 
having different wave polarizations. The transfer function, however,
ignores the nature of the source creating the wave; no averaging
was performed over parameters describing the sources themselves.  Thus, the 
resulting gravitational wave sensitivity curve is equally appropriate 
for most types of sources.  Of course, because of the averaging over 
direction, it is not completely appropriate for a source (like the 
interacting white dwarf binary AM {\it CVn}) whose direction is known.  
Nevertheless, this sensitivity curve seems to us to be a valuable 
intermediate tool to characterize the capability of gravitational wave 
detectors, especially since the strengths of the waves from many 
different types of sources may be plotted on this same graph.  To 
calculate the gravitational-wave strength for a particular type of 
source as it is to be plotted on the sensitivity curve graph requires
some additional steps specific to each class of sources.

\subsection{Sensitivity to Particular Types of Sources}

For {\it continuous monochromatic sources} like circular compact 
binaries, the ultimate source sensitivity comes when the source is 
sampled for a long period of time so as to narrow the bandwidth, 
$\Delta f$.  A continuous source with frequency $f$ and amplitude $h$ 
that is observed over a time $T$ will appear in a Fourier spectrum of 
the data as a single spectral line with root spectral density
\begin{equation}
   h_f = {h\over\sqrt{\Delta f}} = h\sqrt{T}.
   \label{continuous}
\end{equation}
The frequency $f$ and the amplitude $h$ will depend on the several 
parameters of the binary such as total mass, semi-major axis, distance 
to the binary, inclination to the line-of-sight, etc.  If one wants to 
consider a population of monochromatic binaries, then an average over 
the parameters of possible systems should be taken, and the value used 
for $h$ at each frequency should reflect this process.

For signals that are {\it short bursts}, such as would be produced by a 
compact body in highly elliptical orbit about a massive black hole or 
by high-velocity encounters of massive compact bodies in a region of 
high stellar density, the signal would typically have a characteristic 
pulse width $\tau$ with a broad spectral content and would occur only 
once in the lifetime of the detector.  The detection of such a source is 
optimal when the bandwidth $\Delta f$ is not much larger than 
$\tau^{-1}$.  The relationship between the amplitude of the pulse $h$ 
and the strength of the signal as it would be plotted on the root 
spectral density graph is therefore given by 
\begin{equation}
   h_f = {h\over\sqrt{\Delta f}} = h\sqrt{\tau}.
   \label{pulse1}
\end{equation}
Again, if a population of burst sources is to be considered, then one
should average the amplitudes over the possible parameters of the 
source, including all source orientations and accounting for the short 
duty cycles that such sources typically have.

Finally, one may consider how to include a {\it stochastic background} 
of gravitational waves as a source on the gravitational wave 
sensitivity graph.  In some cases, such as the case of a stochastic 
superposition of close compact binary stars, the amplitude spectral 
density $S_{h}$ is known.  In these cases, the root spectral density
\begin{equation}
   h_f(f) = \sqrt{S_h(f)}
   \label{pulse3}
\end{equation}
can be plotted directly on the sensitivity graph.  In other cases, 
what is assumed is a spectrum of the cosmic energy density, 
$\Omega (f)$.  The relationship between amplitude and energy density 
is given by
\begin{equation}
   S_{h}(f) = {{3 H^{2}} \over {\pi^{3} f^{2}}}\Omega(f)
   \label{CosmicSpectralDensity}
\end{equation}
where $H$ is the Hubble constant. The square root of $S_h(f)$ may again
be plotted directly on the sensitivity graph.  Finally, a cosmic 
background is often assumed to 
be simply peaked at some frequency, $f_{p}$, and spread over a decade 
of bandwidth.  In this case, the bandwidth will be approximately equal 
to the peak frequency, and a point plotted at
\begin{equation}
   h_f = {3H\over\pi^{3/2}f_{p}^{3/2}}\Omega^{1/2}
   \label{pulse}
\end{equation} 
would represent the peak of the decade-wide spectrum corresponding to 
a total integrated energy density of $\Omega$.

\subsection{Reconciliation with Other Results from the Literature}

Substantially different methods have often been used in the previous 
literature to illustrate the capability of a spaceborne interferometer 
to detect astrophysical sources of gravitational waves.  This section 
describes how to relate the methods used in this work to the approach 
usually taken in the literature associated with the proposed LISA 
mission \cite{LPPA2}, and the comprehensive development of sensitivity 
limits provided in the review article of Thorne \cite{Thorne300}.

As in this paper, other analyses create an initial sensitivity curve 
for LISA by analyzing the various noise sources in the interferometer 
and then dividing by a transfer function \cite{LPPA2,schilling}.  The 
LISA instrumental sensitivity, in terms of $h_f$, is generally 
converted to an effective gravitational wave amplitude sensitivity, 
$h$, by dividing the value of $h_f$ at each frequency by the square 
root of an assumed one year integration time.  The resulting 
sensitivity curve in $h$ is also multiplied by a factor of $5$, so 
that the final curve indicates a one-year integrated threshold for a 
signal with signal-to-noise ratio $\geq 5$.  The combination of these 
factors is
\begin{equation}
	h_{\rm{SNR} 5}^{1 \rm{yr}} = 8.9 \times 10^{-4} h_f \, \, ,
   \label{LISAtrans}
\end{equation}
where $h_f$ is the spectral amplitude in units of ${\rm Hz}^{-1/2}$.  
This relation assumes that the spurious factor of $\sqrt{3}$ noted 
in Section \ref{sec:Compare} has been corrected.

It should be emphasized that, after having assumed a one-year 
integration time, it is only appropriate to plot monochromatic 
gravitational wave sources against this type of sensitivity curve, and 
only for a case where a signal-to-noise ratio of 5 is actually 
required.  The danger in using such a graph to characterize the 
overall detector sensitivity is that it is subject to easy 
misinterpretation, which has often occurred in the literature.  
Examples include Hogan \cite{hogan}, which overestimates the LISA 
sensitivity to a cosmic stochastic background by over three orders of 
magnitude by misunderstanding how the 1-year integrated curve would 
apply to a stochastic signal, and Aguiar {\it et.\ al.} \cite{aguiar}, 
which significantly overestimates the LISA sensitivity to bursts from 
black hole oscillations lasting a few minutes by using the curves 
representing coherent integration of a signal for a year.\footnote{The 
LISA team has recognized the difficulty of using such plots to 
characterize the detector sensitivity to bursts\cite{LPPApage}.} On 
the other hand, the usual LISA sensitivity curve graph underestimates 
LISA's sensitivity to monochromatic known sources, such as AM $CVn$, 
for which a signal-to-noise ratio of 5 is an excessive requirement.

Another approach is taken in the comprehensive review article by 
Thorne \cite{Thorne300}. In this review, Thorne carefully distinguishes 
between the three types of source, and analyzes the response of a 
detector to each type. For each type of source, he finally derives 
a detector sensitivity which is denoted by $h_{3/yr}(f)$, defined as:
\begin{equation}
	h_{3/yr}(f)  =  11 [f S_h(f)]^{1/2} \,\,\, {\rm burst \ sources} ,
\label{burstsource}
\end{equation}
\begin{equation}		 
	h_{3/yr}(f)  =   3.8[S_h(f) \times 10^{-7} {\rm Hz}]^{1/2} \,\,\,
		{\rm periodic \ sources} , 
\label{periodicsource}
\end{equation}
\begin{equation}
	h_{3/yr}(f)  =  4.5 \left( {\Delta f \over 10^{-7} {\rm Hz}} \right)
		^{-1/4} [f S_h(f)]^{1/2} \,\,\, {\rm stochastic \ background} 
		\ \ . 
\label{stochsource}
\end{equation}
Here $S_h(f)$ is the same spectral density of detector noise utilized 
in this paper.  The derived sensitivity, $h_{3/{\rm yr}}$, represents 
the weakest level at which one has 90\% confidence of detecting a
signal using two cross-correlated identical detectors. For burst 
sources, the level is set by demanding three detections per year. For 
periodic sources, $h_{3/{\rm yr}}$ represents the level obtained in a 
$1/3$ year integration, assuming the frequency and phase of the source 
are known. Finally, for stochastic sources, $h_{3/{\rm yr}}$ represents the 
weakest source that can be detected with a $1/3$ year integration of 
the cross-correlation function between two independent detectors.

The development of Thorne's three dicta for detection of burst, 
periodic, and stochastic sources involves detailed assumptions which 
are described at length in Ref.\ \cite{Thorne300}; that discussion 
will not be reproduced in full here.  It is worth noting, however, 
that not all of the assumptions made in the development of these dicta 
are valid for a space-based interferometer. For example, Thorne assumes 
one has two independent cross-correlated detectors.  While this is a 
valid assumption for LIGO, it is not for a LISA-style instrument.  
Although the data from the three LISA arms can be treated as two 
interferometers, and hence measure both polarizations of gravitational 
waves, the two interferometers are not completely independent, as they 
share an arm.

\section{SUMMARY}\label{sec:Summary}

Constructing the sensitivity curves for a spaceborne observatory, 
such as those shown in Figure \ref{NoiseCurves}, is the first step in 
understanding the response of the instrument to gravitational 
radiation.  The formalism developed here can be used to determine the 
sensitivity of {\it any} space-based interferometer simply in terms of 
the essential parameters which describe the overall design of the 
instrument.  Not only does this allow one to quickly and accurately 
assess the performance of one proposed observatory compared to 
another, but it also provides a quick and easy method for considering 
new observatory designs.  

Carefully detailing the response of any new instrument prepares us for
the inevitable detection of unexplainable signals from distant
astrophysical sources, and provides a clear idea of how to improve our
instrumentation for the construction of the next generation of
observatories.

\acknowledgements
The work of S.\ L.\ L.\ and W.\ A.\ H.\ was was supported in part by National 
Science Foundation Grant No. PHY-9734834 and NASA Cooperative Agreement
No. NCC5-410.

\begin{figure} 
  \caption{ A typical configuration for a spaceborne gravitational 
  wave interferometer.  Three probes form an equilateral triangle, 
  inscribed on the relative orbits of the spacecraft.  A single 
  Michelson interferometer is formed using any two legs of the 
  triangle, and a second (non-independent) interferometer can be 
  formed using the third arm of the configuration in conjunction with 
  an arm already in use for the primary signal.}
  \label{ProbeGeometry}
\end{figure}

\begin{figure} 
   \caption{The root spectral density of the noise in the LISA 
   interferometer formed by adding the strain noises induced by
   acceleration noise and by total position noise in quadrature.}
\label{NoiseSpectrum}
\end{figure}

\begin{figure} 
  \caption{ The geometrical relationship of the interferometer to the 
  propagation vector of a gravitational wave, used to conduct spatial 
  averaging.  The arms are designated by vectors $\bf{L_{1}}$ and 
  $\bf{L_{2}}$ (solid black vectors), while the propagation vector of 
  the gravitational wave is given by $\bf{k}$ (open white arrow).  One 
  arm of the interferometer is aligned along the polar axis of a 
  2-sphere, the other arm lying an angular distance $\gamma$ away long 
  a line of constant longitude.  The angles $\theta_{i}$ relate the 
  vector $\bf{k}$ to the arms of the interferometer, and the angle 
  $\epsilon$ is the inclination of the plane containing $\bf{k}$ and 
  $\bf{L_{1}}$ to the plane of the interferometer.}
  \label{GeometryFig}
\end{figure}

\begin{figure} 
  \caption{ The transfer function $R(u)$ is shown as a function of the 
  dimensionless variable $u = \omega \tau$.  Note that it is roughly 
  constant at low frequencies, and has a ``knee'' located at $u = 
  \omega \tau \sim 1$.}
  \label{TransferPlot}
\end{figure}

\begin{figure} 
  \caption{ The sensitivity curves for the proposed LISA observatory 
  is shown.  The low frequency rise is due to acceleration noise in 
  each of the systems.  The high frequency rise is due to the ``knee'' 
  in the transfer function at $f \simeq (2 \pi \tau)^{-1}$.  The 
  structure at high frequencies is a consequence of the high frequency
  structure in the gravitational wave transfer function.}
  \label{NoiseCurves}
\end{figure}

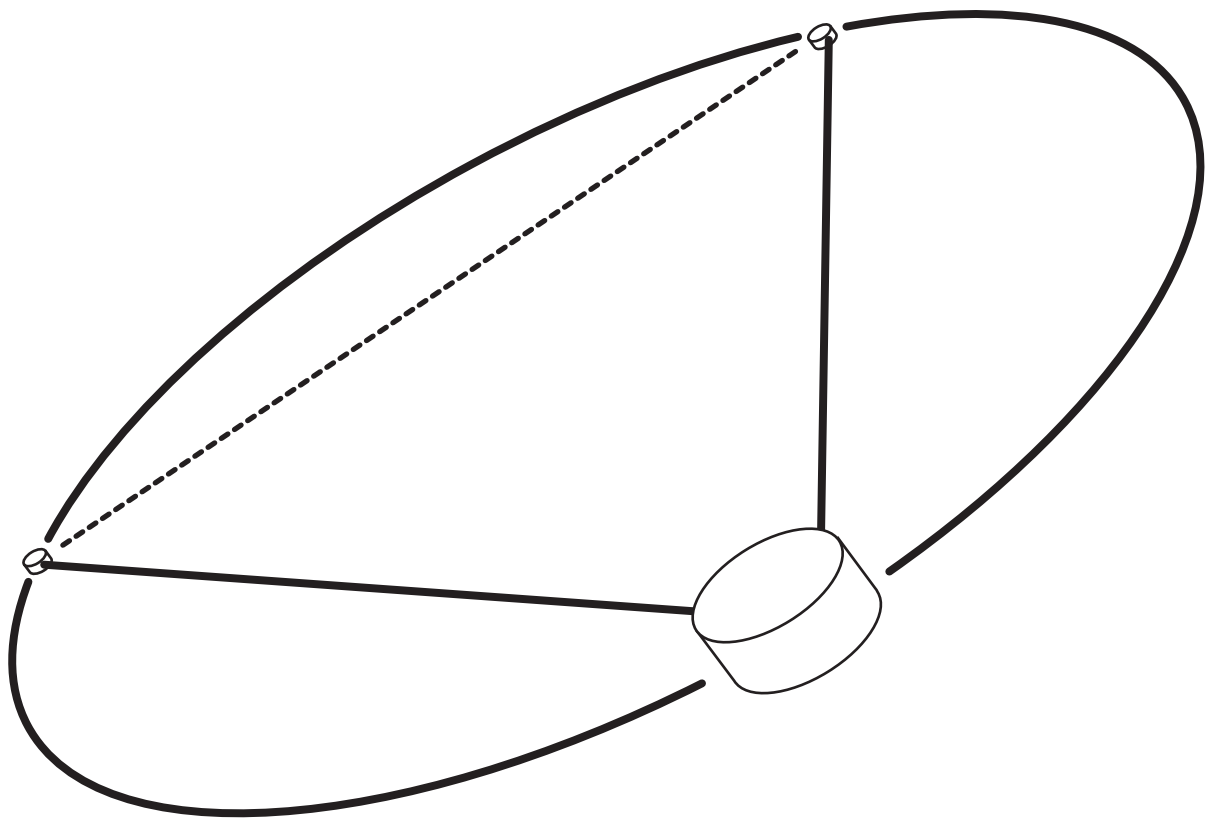
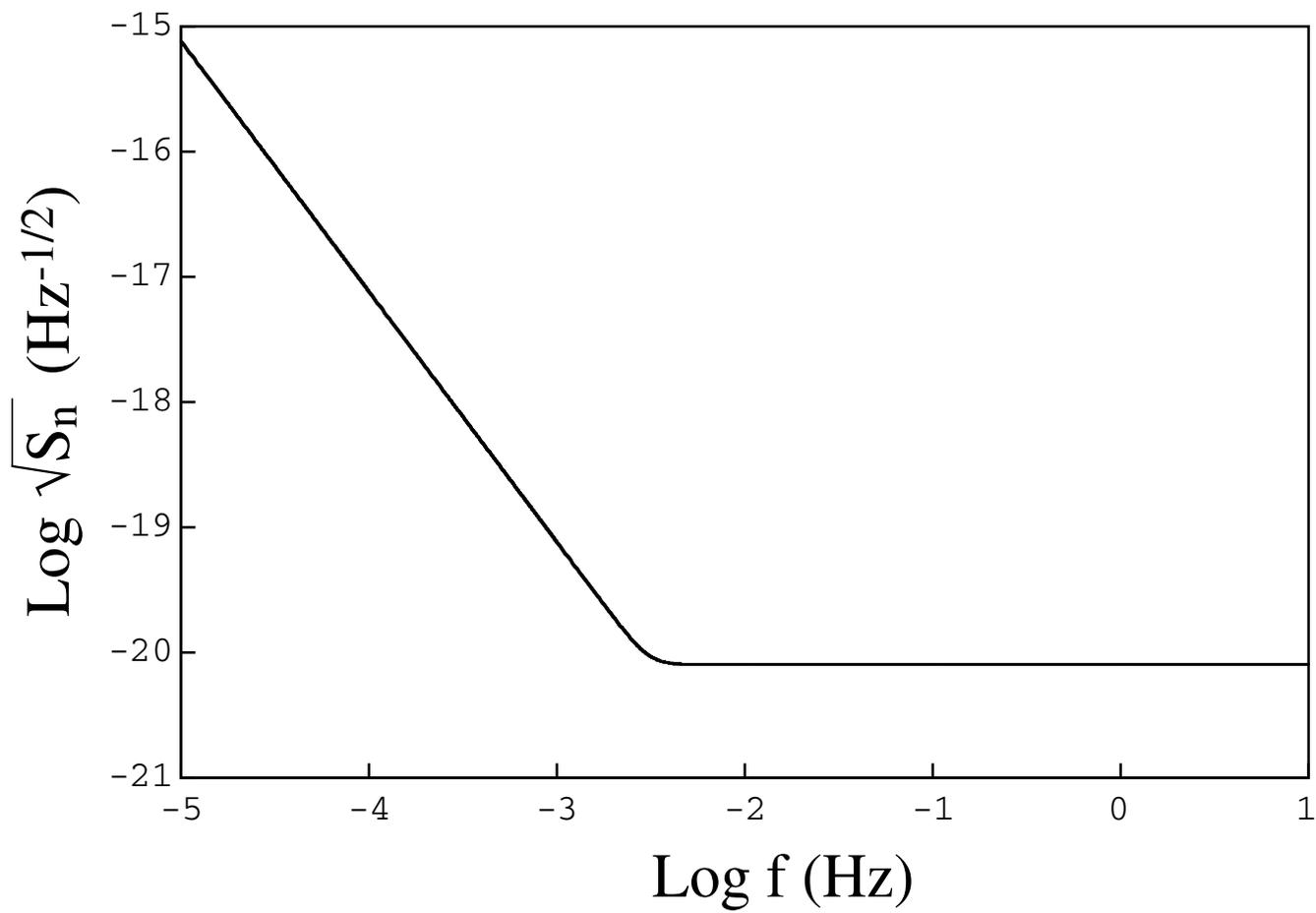
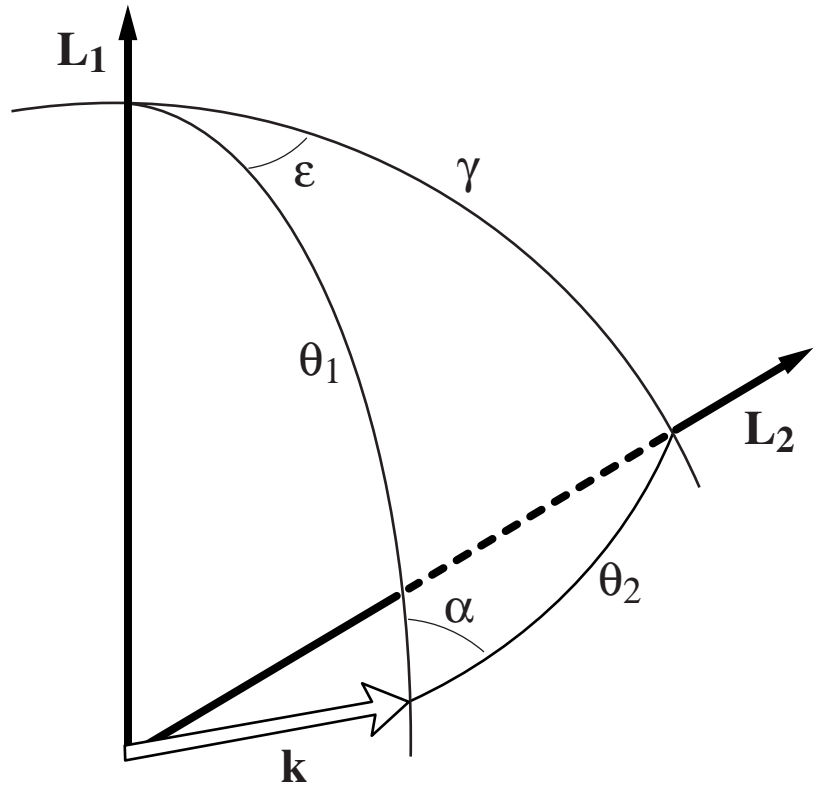
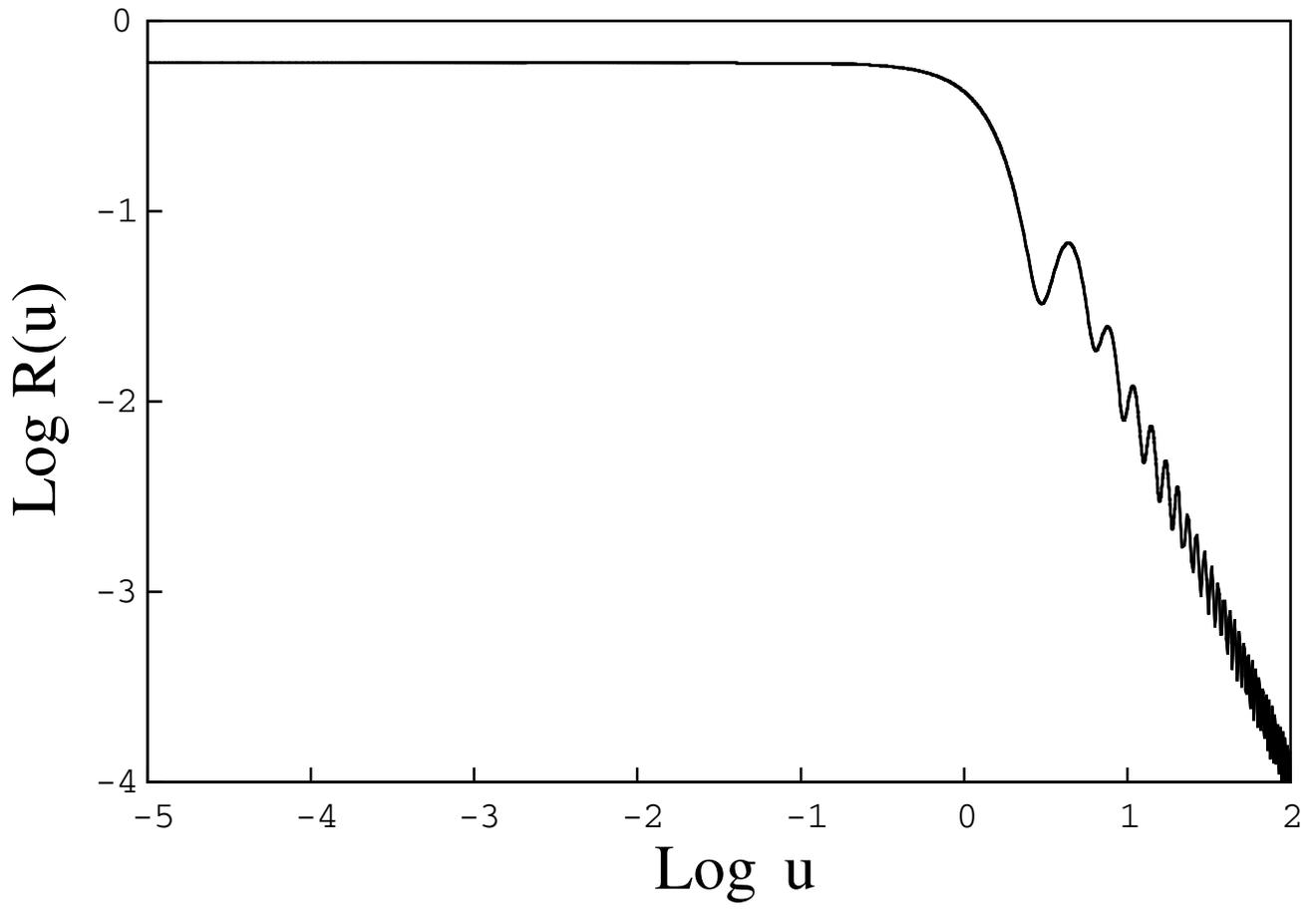
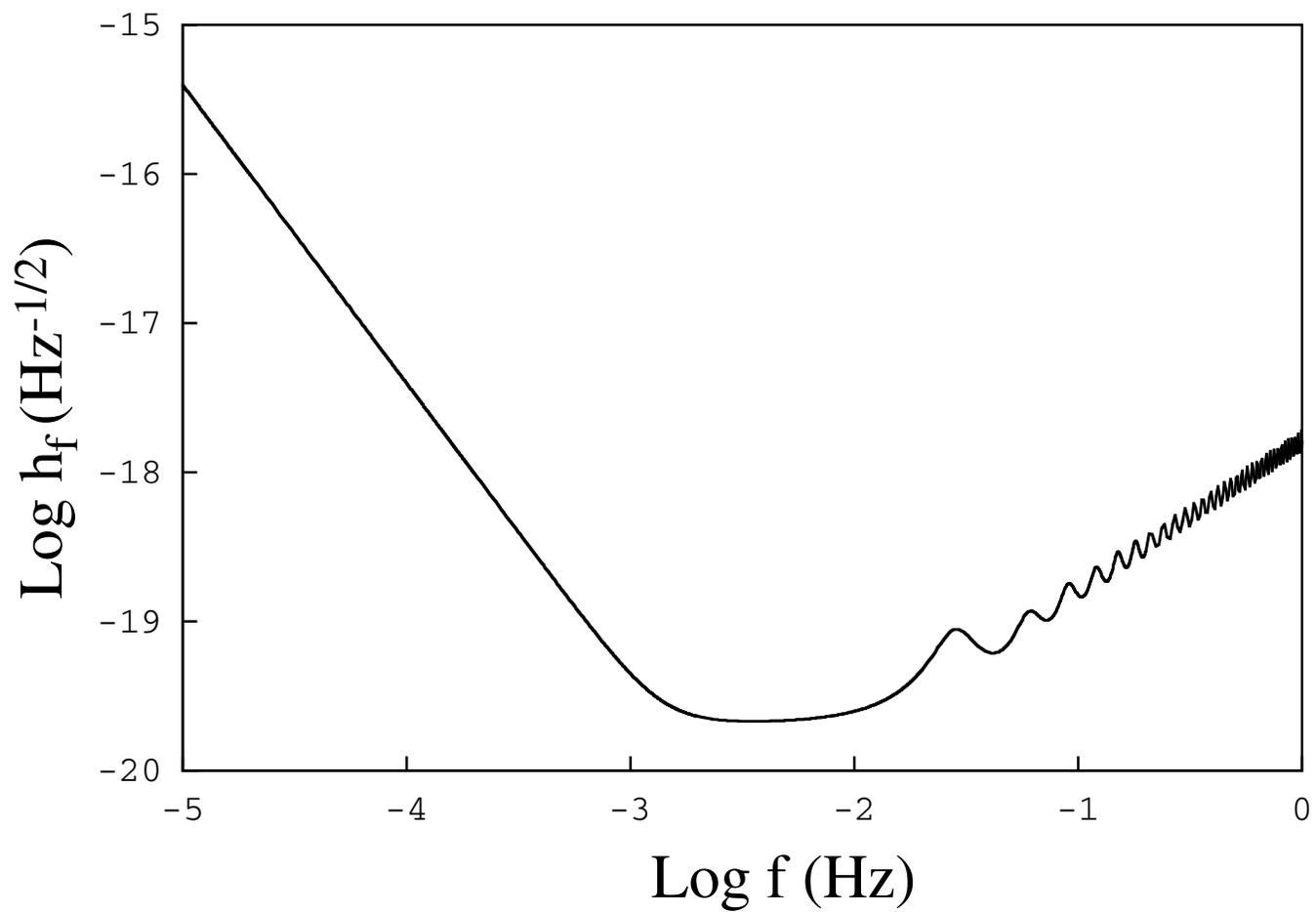
\end{document}